\documentclass[prd,article ]{revtex4}%
\usepackage{caption}
\usepackage{amsmath}
\usepackage{graphicx}
\usepackage{epsfig}
\usepackage{dcolumn}
\usepackage{bm}
\usepackage{slashed}
\usepackage{amsfonts}
\usepackage{amssymb}
\usepackage{float}%
\providecommand{\U}[1]{\protect\rule{.1in}{.1in}}
\begin{document}
\title{A QCD Lagrangian including renormalizable NJL terms. }
\author{Alejandro Cabo Montes de Oca $^{a,b}$,}
\address{ $^{a}$ Max-Planck-Institut fur Physik (Werner-Heisenberg-Institut),\\
Fohringer Ring 6, D-80805 Munich, Germany,\\
$^{b}$ Departamento de F\'{\i}sica Te\'{o}rica, Instituto de
Cibern\'{e}tica,Matem\'atica y F\'isica,\\ Calle E, no. 309, Vedado, La Habana,
Cuba.
\bigskip
}
\begin{abstract}
\noindent A local and gauge invariant version of  QCD Lagrangian is introduced.
The model includes Nambu-Jona-Lasinio (NJL) terms within its action in a
surprisingly renormalizable form. This occurs thanks to the presence of action
terms which at first sight, look as breaking power counting renormalizability.
However, those terms also modify the quark propagators, to become more
decreasing that the Dirac one at large momenta, indicating power counting
renormalizability. The approach, can also be interpreted as generalized
renormalization procedure for massless QCD. The free propagator, given by the
substraction between a massive and a massless Dirac ones, in the Lee-Wick
form, suggests that the theory also retains unitarity. The appearance of
finite quark masses already in the tree approximation in the scheme is
determined by the fact that the new action terms explicitly break chiral
invariance. The approach looks as being able to implement the Fritzsch
Democratic Symmetry breaking ideas about the quark mass hierarchy. Also, a
link of the theory with the SM seems that can follow after employing the
Zimmermann's couplings reduction scheme. The renormalized Feynman diagram
expansion of the model is written here and the formula for the degree of
divergence of the diagrams is derived. The primitive divergent graphs are identified.
We start their evaluation through the calculation of the
divergence of the gluon polarization operator. The divergent  part of the gluon selfenergy
results to be transverse as required by gauge invariance and also independent of the new couplings
 appearing. The calculations of the  rest of primitive divergences will be presented elsewhere.

\end{abstract}
\maketitle

\address{ $^{a}$ Max-Planck-Institut für Physik (Werner-Heisenberg-Institut),\\
Föhringer Ring 6, D-80805 Munich, Germany,\\
$^{b}$ Departamento de F\'{\i}sica Te\'{o}rica, Instituto de
Cibern\'{e}tica,Matem\'atica y F\'isica,\\ Calle E, no. 309, Vedado, La Habana,
Cuba.
\bigskip
}

\section{Introduction}

Determining the origin of the wide range of values spanned by the quark
masses, and more generally, the structure of the lepton and quark mass
spectrum, is one of the central problems of High Energy Physics. We have
considered a previous search associated to this question. It was motivated by
the suspicion about that the large degeneration of the non-interacting
massless QCD vacuum (the state which is employed for the construction of the
standard Feynman rules of PQCD) in combination with the strong forces carried
by the QCD fields, could be able to generate a large dimensional transmutation
effect. This fact, in turns, could then imply the generation of quark and
gluon condensates and masses. The investigation of quark and gluon
condensation effects had been widely considered in the literature
\cite{1,2,3,4,5,6,7,8}. Our previous works on the theme appeared in references
\cite{9,10,11,12,13,14,15,16,17}. Assumed that the idea in them is correct,
the following picture could arise. A sort of Top condensate model might be the
effective action for massless QCD. In it, a Top quark condensate, arising
within the same inner context of the SM, could play the role of the Higgs
field. Thus, the SM could be \textquotedblleft closed\textquotedblright\ by
generating all the masses within its proper context. We imagine this effect
might occurs as follows: In a first step, the six quarks could get their
masses thanks to a flavour symmetry breaking determined by the quark and gluon
condensates. Afterwards, the electron, muon and tau leptons, would receive
their intermediate masses thanks to radiative corrections mediated by the mid
strength electromagnetic interactions with quarks. Finally, the only weak
interacting character of the three neutrinos with all the particles, could
determine their even smaller mass values.

Its is clear, that such a picture will need to satisfy the  strong
constraints imposed by the large experimental evidence about the
validity of the SM model, which had came from the LHC. Specifically, some current evaluations of
energy scales being  compatible with the experimental information,
and with  Higgs compositeness are in the range of   $500$ GeV \cite {higgscom-1,higgscom-2}.
 Therefore, since  the above energy bound is not extremely high, we have no enough
appealing reasons to discard the indications at hand about
that  the idea investigated in this work can be helpful. Let us resume briefly those
indications. In Ref. \cite{10}, with the use of a BCS squeezed state like
vacuum state (formed with nearly zero momenta gluons and ghost particles)
modified Feynman rules for massless QCD were derived. The case of gluon
condensation in the absence of quark pair condensation was initially
considered. Then, a proper selection of the parameters allowed to derive an
addition to the gluon free propagator: a Dirac's Delta function centered at
zero momenta multiplied by the metric tensor. Such a term were before
discussed by Munczek and Nemirovsky in \cite{2}. Before, in reference \cite{9}
it was simply proposed this modification and used to argue that it predicts a
non vanishing value of the gluon condensate in the first corrections. The
physical state and zero ghost number conditions were also imposed to fix the
parameters of the squeezed vacuum. Then, the results obtained for gluons
motivated the idea of also considering the quarks as massless and to search
for the possibility to generate their masses dynamically, thanks to the
condensation of quark pairs. For this purpose the BCS like initial state was
generalized in reference \cite{11} to include the quark pair condensates in
massless QCD. In this case, in a similar way as for gluons, the quark
propagators simply were modified again by the addition of a term being the
product of a Dirac's Delta function at zero momentum and the spinor identity
matrix. Next, in Refs. \cite{11,12} the main conclusion obtained from this
starting approach followed from a simple discussion of the Dyson equation for
quarks. It was considered by taking the quark self-energy in its lowest order
in the power expansion in the condensate parameters. The coefficient of the
zero momentum Delta function was fixed to reproduce the estimate of the
gluonic Lagrangian mean value, following from the sum rules approaches. After
that, the solution of the Dyson equation was able to predict the
\textquotedblleft constituent\textquotedblright values of 1/3 of the nucleon
mass for the light quarks. The initial approach was also further studied in
\cite{13,14} order to define a regularization scheme for eliminating the
singularities which could appear in the Feynman diagram expansion due to the
Delta functions at zero momentum entering in the modified propagators.
However, due to the unusual characteristics of the approach, we decided in
reference \cite{15,16} to also investigate the possibility of re-expressing
the condensation effects in the modified propagators as equivalent vertices in
the Lagrangian.

The result of this work was the central step in leading to the model to be
presented here. It was obtained that the condensate effects introduced in
massless QCD by employing a squeezed state as modified vacuum, were equivalent
to the addition of a new four legs vertex term in the Lagrangian, including
two gluon and two quark lines. But, the vertex was a non local one including a
zero momentum delta function. The obtained vertex term structure was not a
fully gauge invariant one, but this drawback can be understood as due to the
simple non gauge invariant form employed for the squeezed free vacuum state.
However, the resulting curious structure directly led to the idea of
constructing a local and gauge invariant form of the theory: It became clear
that it is possible to include a similar kind of two gluon and two quark
vertices, presumably incorporating the gluon and quark condensates, but in a
gauge invariant and local form. This modification was presented in reference
\cite{17}, where, it was also possible to argue that the new terms added to
the action do not break the power counting renormalizability of massless QCD.
This observation, then led to a surprising conclusion, also exposed in
reference \cite{17}, about that the Nambu-Jona-Lasinio (NJL) four fermion
vertices also turn out to become renormalizable counterterms of the considered
Lagrangian. The resulting theory included an additional set of six fermion
fields showing a negative metric. However, the modified quark propagator also
showed a rapidly\ decaying momentum dependence thanks to its Lee-Wick
structure \cite{leewick1} which suggests that the negative metric states could
result to be non propagating thanks to the radiative corrections
\cite{leewick2}. The mentioned properties opened the opportunity that the
proposed model can show the mass generation properties which the NJL models exhibit.

In the present work we start properly defining the model proposal and its
consequences. That is, to begin we will concretely define the local and gauge
invariant form of QCD Lagrangian being power counting renormalizable. The
action will simply be the massless QCD one, plus six additional terms, one for
each flavour, of similar vertices formed by products of two quark and two
covariant derivatives. The Lagrangian will also include new NJL type of four
quark vertices, which are not usually allowed by the power counting
renormalizability. Further the free gluon and quark propagators will be
determined by evidencing the modified Lee-Wick structure of the quark
propagators quadratically decaying at large momenta. Then, the formula for the
divergence indices of the Feynman diagrams will be derived, allowing to
evidence the possibility of adding the NJL terms in a power counting
renormalizable form. Next, the renormalized form of the action will be written
and the expressions and diagrams for all the usual and counterterm vertices
defined. The renormalized diagram expansion is then employed to evaluate the
one loop divergences of the gluon selfenergy. The divergence of the
polarization tensor, as imposed by the gauge invariance, resulted to be
transversal. The calculation of the rest of the large number of primitive
divergences will be presented elsewhere.

Section II presents the Lagrangian structure of the model and writes its
associated propagators and vertices to discuss the power counting
renormalizability. Next, in Section III the index of divergence is derived and
the primitive divergences are identified. Section IV presents in detail the
renormalized Feynman expansion. The resulting perturbative expansion is
employed in Section V for evaluating the one loop divergences of the gluon
selfenergy. The results are summarized in the last section.

\section{The QCD Lagrangian including NJL terms}

The action of the model in the extended $D$ dimensional Minkowsky space is
written in the local and gauge invariant form%
\begin{align}
S &  =\int dx^{D}\text{ }{\Large (}\,-\frac{1}{4}F_{\mu\nu}^{a}(x)F^{a\mu\nu
}(x)-\frac{1}{2\alpha}\partial_{\mu}A^{a\mu}(x)\partial_{\nu}A^{a\nu
}(x)+\partial_{\mu}\chi^{\ast a}(x)D^{ab\mu}\chi^{b}(x)+\nonumber\\
&  +\sigma\sum_{q}\overline{\Psi}_{q}^{i}(x)\text{ }i\gamma^{\mu}D_{\mu}%
^{ij}\Psi_{q}^{j}(x)-\sum_{q}\varkappa_{q}\,\overline{\Psi}_{q}^{j}\text{
}(x)\overleftarrow{D}^{ji\mu}\text{ }D_{\mu}^{ik}\Psi_{q}^{k}%
(x)\,+\label{action}\\
&  +\sum_{f}\sum_{q_{1},q_{2},q_{_{3}},q_{_{4}}}\frac{\lambda_{f}\text{ }}%
{4}^{(f)}\Lambda_{(j_{2},r_{2},q_{_{2}})(j_{4},r_{4},q_{4})}^{(j_{1}%
,r_{1},q_{1})(j_{3,},r_{3},q_{3})}\text{ }\overline{\Psi}_{q_{_{1}}}%
^{j_{1},r_{1}}(x)\overline{\Psi}_{q_{_{2}}}^{j_{2},r_{2}}(x)\Psi_{q_{_{3}}%
}^{j_{3},r_{3}}(x)\Psi_{q_{4}}^{j_{4},r_{4}}\text{ }(x){\LARGE )},
\end{align}
where $i,k,j,...=1,2,3$ are color indices and the spinor ones are hidden to
simplify notation, the $q=1,...,6$ indices indicate the $flavour$ of the
quarks. It should be underlined that the main different elements in this
action with respect to the massless QCD, are the presence of the two last
terms and the possible change in the sign of the Dirac Lagrangian implied by
the values considered for the constant $\sigma=\pm1$. Note the appearance of
six new couplings $\varkappa_{q}$, one for each quark (flavour) index $q.$ The
last term is the added Nambu-Jona-Lasinio like four quarks action. The
coefficients $^{(f)}\Lambda_{(j_{2},r_{2},q_{_{2}})(j_{4},r_{4},q_{4}%
)}^{(j_{1},r_{1},q_{1})(j_{3,},r_{3},q_{3})}$ are assumed to be such that the
corresponding action term in the above Lagrangian remains invariant under an
arbitrary symmetry transformation of the quark fields (color, $flavour$ or
Lorentz ones) for each particular of the let say $F$ values of the index
$\ f=1,2,...,F$ . For bookkeeping purposes, the conventions for the various
quantities are defined as follows
\begin{align}
F_{\mu\nu}^{a} &  =\partial_{\mu}A_{\nu}^{a}-\partial_{\nu}A_{\mu}^{a}-g\text{
}f^{abc}A_{\mu}^{b}A_{\nu}^{c},\\
\Psi_{q}^{k}(x) &  \equiv\left(
\begin{array}
[c]{c}%
\Psi_{q}^{k,1}(x)\\
\Psi_{q}^{k,2}(x)\\
\Psi_{q}^{k,3}(x)\\
\Psi_{q}^{k,4}(x)
\end{array}
\right)  ,\\
\Psi_{q}^{\dag k}(x) &  \equiv(\Psi_{q}^{k}(x))^{T\ast}\nonumber\\
&  =\left(
\begin{array}
[c]{cccc}%
(\Psi_{q}^{k,1}(x))^{\ast} & (\Psi_{q}^{k,2}(x))^{\ast} & (\Psi_{q}%
^{k,3}(x))^{\ast} & (\Psi_{q}^{k,4}(x))^{\ast}%
\end{array}
\right)  ,
\end{align}
where, as mentioned before, $q$ $=1,...,6$ indicates the $flavour$ index. The
expressions for the Dirac conjugate spinors and covariant derivatives are
\begin{align}
\overline{\Psi}_{q}^{j}\text{ }(x) &  =\Psi_{q}^{\dag k}(x)\gamma^{0},\text{
}\\
D_{\mu}^{ij} &  =\partial_{\mu}\delta^{ij}-i\text{ }g\text{ }A_{\mu}^{a}%
T_{a}^{ij},\ \overleftarrow{D}_{\mu}^{ij}=-\overleftarrow{\partial}_{\mu
}\delta^{ij}-i\text{ }g\text{ }A_{\mu}^{a}T_{a}^{ij},\\
D_{\mu}^{ab} &  =\partial_{\mu}\delta^{ab}-g\text{ }f^{abc}\text{ }A_{\mu}%
^{c},
\end{align}
in which the Dirac's matrices, $SU(3)$ generators and the metric tensor are
defined in this section in the conventions of reference \cite{muta}, as
\begin{align}
\{\gamma^{\mu},\gamma^{\nu}\} &  =2g^{\mu\nu},\ \ \ \ [T_{a},T_{b}]=i\text{
}f^{abc}T_{c}.\text{ }\gamma^{0}=\beta,\text{ }\gamma^{j}=\beta\text{ }%
\alpha^{j},\text{ }j=1,2,3,\nonumber\\
g^{\mu\nu} &  \equiv\left(
\begin{array}
[c]{cccc}%
1 & 0 & 0 & 0\\
0 & -1 & 0 & 0\\
0 & 0 & -1 & 0\\
0 & 0 & 0 & 1
\end{array}
\right)  ,\text{ }\beta=\left(
\begin{array}
[c]{cc}%
I & 0\\
0 & -I
\end{array}
\right)  ,\text{ }\alpha^{j}\equiv\left(
\begin{array}
[c]{cc}%
0 & \sigma^{j}\\
\sigma^{j} & 0
\end{array}
\right)  ,\text{ }j=1,2,3,\nonumber\\
\sigma^{1} &  =\left(
\begin{array}
[c]{cc}%
0 & 1\\
1 & 0
\end{array}
\right)  ,\text{ }\sigma^{2}=\left(
\begin{array}
[c]{cc}%
0 & -i\\
i & 0
\end{array}
\right)  ,\text{ }\sigma^{3}=\left(
\begin{array}
[c]{cc}%
1 & 0\\
0 & -1
\end{array}
\right)  ,\text{ }I=\left(
\begin{array}
[c]{cc}%
1 & 0\\
0 & 1
\end{array}
\right)  .
\end{align}
Other definitions and relations for the coordinates are
\begin{equation}
x\equiv x^{\mu}=(x^{0},\overrightarrow{x})=(x^{0},x^{1},x^{2},x^{3}),\text{
\ }x_{\mu}=g_{\mu\nu}x^{\nu},\text{ }x^{0}=t.
\end{equation}
The notations to be employed in this work are chosen to coincide with the ones
used in the text in reference \cite{muta}. This criterium was adopted in order
to become able of employing the large set of evaluations, definitions and
auxiliary formulae presented in that reference for QCD. As underlined before,
a basic new element in the proposed action are the six vertices of the form
\begin{equation}
-\sum_{q}\varkappa_{q}\,\overline{\Psi}_{q}^{j}(x)\gamma_{\mu}%
\overleftarrow{D}^{ji\mu}\gamma_{\nu}D^{ik\nu}\Psi_{q}^{k}(x),
\end{equation}
where the six coefficients $\varkappa_{q}$ will be called \textquotedblleft
condensate parameters\textquotedblright, since they enter in similar
`positions' to the parameters appearing in the \textquotedblleft motivating
\textquotedblright non local vertex derived in the previous work \cite{16}.
The $\varkappa_{q}$ are now six new dimensional constants of the theory. A new
element with respect to the massless QCD in the proposed action is allowing a
change in the sign of the Dirac Lagrangian. The interest of this change was
discussed in reference \cite{18}. If the modification in the sign is allowed,
it will lead to free quark propagators which are expressed as usual positive
metric Dirac propagator of massive fermions plus a negative metric massless
propagators. The usual sign assignment determines that the massive propagator
shows negative metric. Since experiments seem to indicate that the massive
quarks in QCD should be physically relevant within the model, the negative
sign of the Dirac action was adopted. However, it can be remarked that in
reference \cite{18} it was also argued that such a change in the sign can be
introduced in the same physical action by a change of field and coordinates
transformation. Thus, since the same action is associated to the quantization
procedure, the fixed negative signs can be associated to a quantization of the
same physical system but using alternative transformed field variables and coordinates.

\subsection{The free propagators}

The Lagrangian associated to the action in (\ref{action}) can be decomposed in
the quadratic in the fields approximation $\mathcal{L}_{0}$ and the one
determining the interaction vertices $\mathcal{L}_{i}$ to write
\begin{align}
S  &  =\int dx^{D}\mathcal{L=}\text{ }S_{0}+S_{i}\nonumber\\
&  =\int dx^{D}\mathcal{L}_{0}+\int dx^{D}\mathcal{L}_{i}.
\end{align}
After finding the inverse of the kernels defining the free form of the action%
\begin{align}
S_{0}  &  =\int dx^{D}(\frac{1}{2}A^{a\mu}(g_{\mu\nu}\partial^{2}-(1-\frac
{1}{\alpha})\partial_{\mu}\partial_{\nu})A^{a\nu}+\nonumber\\
&  -\chi^{\ast a}\partial^{2}\chi^{a}+\sum_{q}\overline{\Psi}_{q}^{i}%
(\sigma\text{ }i\text{ }\gamma^{\mu}\partial_{\mu}+x_{f}\text{ }\partial
^{2})\Psi_{q}^{i}\text{ }),
\end{align}
it follows that the gluon quark and ghost propagators can be evaluated in the
form
\begin{align}
D_{\mu\nu}^{ab}(p)  &  =\frac{\delta^{ab}d_{\mu\nu}(p)}{p^{2}+i\epsilon}%
=\frac{\delta^{ab}(g_{\mu\nu}-(1-\alpha)\frac{p_{\mu}p_{\nu}}{p^{2}})}%
{p^{2}+i\epsilon},\\
S_{q}^{ij}(p)  &  =\frac{\sigma\text{ }\delta^{ij}}{\sigma\text{ }%
\varkappa_{q}\text{ }p^{2}-\gamma_{\nu}p^{\nu}}\nonumber\\
&  =\sigma\text{ }\delta^{ij}(\frac{1}{\gamma_{\nu}p^{\nu}-\frac{1}%
{\sigma\varkappa_{q}}}-\frac{1}{\gamma_{\nu}p^{\nu}}),\text{ }%
q=1,2,...,6,\label{propa}\\
D^{ab}(p)  &  =-\frac{\delta^{ab}}{p^{2}+i\epsilon}=-\frac{\delta^{ab}}%
{p^{2}+i\epsilon}.
\end{align}
The difference between these Green functions and the ones related with
massless QCD is present only in the quark propagators which each of them
appears expressed as a difference between the usual massive Dirac propagator
and one also usual but massless. Note that in dependence of the sign $\sigma$
of the Dirac action, the massive component shows a normal positive or negative
metric. As noted before, the negative sign of $\sigma$ fix the massive term as
showing positive metric. The gluon and ghost free propagators are the usual
ones, and their notation coincides with the one in reference \cite{muta}. The
most curious property of these free propagators is that all of them behave as
$\frac{1}{p2}$ at large momenta. Therefore, since the maximal number of fields
plus derivative factors in any of the Lagrangian terms is four, interestingly,
the model appear to be power counting renormalizable. The masses of each of
the massive quarks become just proportional to the inverse of its
corresponding condensate parameter (couplings) $\varkappa_{q},$ $\ q=1,...,6$.
As it was mentioned, the massive propagator has the appropriate sign
corresponding to positive norm states when $\sigma=-1$. On another hand, the
massless component has the sign related with negative norm states. It can be
remarked that within QCD, assumed to describe Nature, it is currently
interpreted that nor gluons or quarks show asymptotic states. Thus, the
negative metric of the massless free states seem not be a direct drawback of
the model. However, the fact that in very high energy processes, a description
based in massive quarks in short living asymptotic states, seems to describe
the experiences, suggests that an approach in which the massive quarks have
positive norms and massless do not appear (thanks to radiative corrections)
would be convenient. Then, the negative value of $\sigma$ was allowed to be
considered in the model.

\subsection{The vertices}

\begin{figure}[h]
\begin{centering}
\center
\includegraphics[scale=0.2]{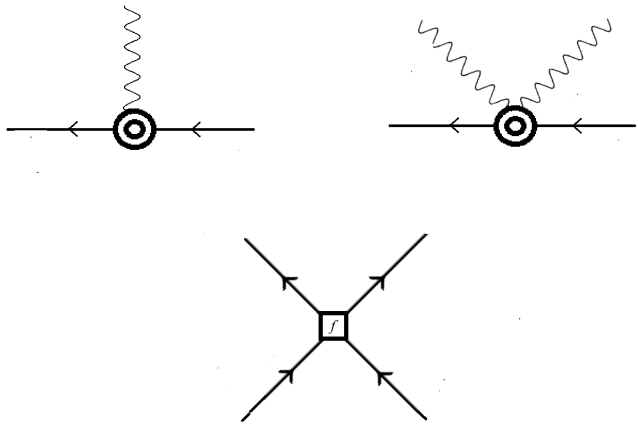}
\par\end{centering}
\caption{The new types of vertices. The three legs ones, at difference with
the usual thee legs vertices in usual QCD, does not contains linear terms in
the gamma matrices. The two-gluon-two-quark four legs vertices are the local
counterparts of the non local vertices representing quark condensate effects
obtained in references \cite{15,16}. The four quark legs vertices are the ones
generated by the NJL four quark interactions. There is a maximal number $F$ of
such vertices being invariant under the symmetries of QCD. }%
\label{figure1}%
\end{figure}

The modified action (\ref{action}) determines new vertices in addition to the
usual in massless QCD. The new types of vertices appearing are illustrated in
figure \ref{figure1}. There are a total of $16+F$ different vertices: $v_{i}%
,i=1,...16+F$. The $\ v_{1}$ to $v_{4}$ will indicate the usual ones in
massless QCD, the three gluon legs, four gluon legs and the ghost-gluon and
quark-gluon interaction vertices, respectively. The additional $v_{5}$
$-v_{10}$ vertices are associated to the new six kinds of three legs quark
gluon interaction. Further, the $v_{11}-v_{16}$ additional new vertices
correspond to the special new 4 legs vertices in which two gluon and two
quarks interact. It should be recalled that precisely these the new added
vertices were motivating the interest in the model under consideration.
Finally, the last new vertices $v_{17}-v_{16+F}$ are associated with the now
allowed $F$ of the NJL type of terms included in the Lagrangian, where $F$ is
the number introduced $\Lambda$ interaction terms defining the four fermion
actions which remain invariant under the symmetries of QCD. The interaction
Lagrangian terms corresponding to these mentioned vertices have the forms%
\begin{align}
\mathcal{L}_{i} &  =\frac{g}{2}f^{abc}(\partial_{\mu}A_{\nu}^{a}-\partial
_{\nu}A_{\mu}^{a})A^{b\mu}A^{c\nu}-\frac{g^{2}}{4}f^{abe}f^{cde}A_{\mu}%
^{a}A_{\nu}^{b}A^{c\mu}A^{d\nu}-\nonumber\\
&  -g\text{ }f^{abc}(\partial^{\mu}\chi^{a\ast})\chi^{b}A_{\mu}^{c}%
+\sum_{q=1,...,6}\sigma g\overline{\Psi}_{q}^{i}\gamma^{\mu}T_{a}^{ij}\Psi
_{q}^{j}A_{\mu}^{a}-\nonumber\\
&  -i\sum_{q=1,...,6}\varkappa_{q}\text{ }g\text{ }\left(  \overline{\Psi}%
_{q}^{i}\partial^{\mu}(T_{a}^{ij}\Psi_{q}^{j}A_{\mu}^{a})+\overline{\Psi}%
_{q}^{i}A_{\mu}^{a}\partial^{\mu}(T_{a}^{ij}\Psi_{q}^{j})\right)  +\nonumber\\
&  +\sum_{q=1,...,6}\text{ }\varkappa_{q}\text{ }g^{2}\text{ }\overline{\Psi
}_{q}^{i}A^{c\mu}T_{c}^{ij}T_{b}^{jk}A_{\mu}^{b}\Psi_{q}^{k}+\nonumber\\
+ &  \sum_{f}\sum_{q_{1},q_{2},q_{_{3}},q_{_{4}}}\frac{\lambda_{f}}{4}\text{
}^{(f)}\Lambda_{(j_{2},r_{2},q_{_{2}})(j_{4},r_{4},q_{4})}^{(j_{1},r_{1}%
,q_{1})(j_{3,},r_{3},q_{3})}\text{ }\overline{\Psi}_{q_{_{1}}}^{j_{1},r_{1}%
}(x)\overline{\Psi}_{q_{_{2}}}^{j_{2},r_{2}}(x)\Psi_{q_{_{3}}}^{j_{3},r_{3}%
}(x)\Psi_{q_{4}}^{j_{4},r_{4}}\text{ }(x).
\end{align}

\section{The divergence index of the model}

Let us consider an arbitrary connected Feynman diagram of the proposed model
including $n_{B}=n_{g}+n_{gh}$ boson internal lines, in which the entering
$n_{gh}$ ghost lines are considered as boson ones as well as the number of
gluon lines $n_{g}.$ The external boson lines are equally constituted as the
sum of the gluon and ghost ones as $N_{B}=N_{g}+N_{gh}.$ Similarly,the number
of internal fermion lines of any type of quark will be denoted as $n_{F}$ and
the corresponding external ones as $N_{F}$\,\,\cite{muta}. For each kind $i$
of the considered vertices of the model, let us define the number of boson
lines connected to it as $b_{i}$, and the number of fermion lines also
attached to it as $f_{i}.$ Define also the total number of vertices of kind
$i$ appearing in the diagram as $n_{i}$ and the number of spatial derivatives
entering in the definition of the vertex $\delta_{i}$ . Therefore, the total
number of derivatives which appear in the diagram $\delta$ can be written as
\begin{equation}
\delta=\sum_{i=1,...,N_{V}}n_{i}\delta_{i}\text{ }.
\end{equation}
But, the total number of free propagator ends associated to the internal boson
lines, plus a half of the propagator ends related to external lines boson
lines should coincide with the total number of boson lines arriving to all the
vertices. Also , the same property is valid for fermion propagators. Thus, the
following relations for any connected (or disconnected) diagram are valid
\begin{align}
2n_{B}+N_{B}  &  =\sum_{i=1,...,N_{V}}n_{i}b_{i},\\
2n_{F}+N_{F}  &  =\sum_{i=1,...,N_{V}}n_{i}f_{i}.
\end{align}
They allow to express the total numbers of boson and fermion lines in terms of
the number of external ones and the parameters defining the vertices as
follows
\begin{align}
n_{B}  &  =\frac{1}{2}(\sum_{i=1,...,N_{V}}n_{i}b_{i}-N_{B}),\\
n_{F}  &  =\frac{1}{2}(\sum_{i=1,...,N_{V}}n_{i}f_{i}-N_{F}).
\end{align}
Let us define for later use, the number of vertices of the interaction of one
gluon with two quarks as defined by the Dirac Lagrangian as $n_{i}$ for $i=4$,
that is as $n_{4}$ $\equiv$ $n_{gqq}$. Considering that the number of
independent $D$ dimensional momentum integrations of the diagram is equal to
the number of propagators minus the number of independent momentum
conservation laws associated to all the vertices, and remembering that this
number is given by the number of vertices minus one for connected diagrams, it
follows for the number of closed loops $l$ in those connected diagrams (the
number of independent momentum integrations) the expression
\begin{equation}
l=\frac{1}{2}\sum_{i=1,...,N_{V}}n_{i}(b_{i}+f_{i}-2)-\frac{1}{2}(N_{B}%
+N_{F})+1.
\end{equation}
Now is possible to define the superficial degree of divergence index of the
diagrams in the model by counting the momenta dependence as
\begin{equation}
d_{G}=D\text{ }l+(\delta-\frac{N_{FP}}{2})-2n_{B}-2n_{F}\text{ },
\label{degree}%
\end{equation}
where $D$ $l$ is the number of independent momentum integrations in the
diagram, $\delta$ is the number of momenta entering in all the definitions of
the appearing vertices, $2n_{B}$ is the number of momenta defining the
decreasing asymptotic behavior of the product of all bosonic propagators
(gluon and ghost ones). Here is important to note that for ghosts, the number
of the momenta appearing in the vertices are not determining an equal number
of divergent momenta factors. This is because, in the number $N_{FP}$ of
ghost-gluon vertices associated to the external ghost lines, only half of them
determines a momentum really diverging when all the independent integration
momenta go to infinity. Therefore the number of divergent momenta determined
by the vertices is reduced to $\delta-\frac{N_{FP}}{2}$\,\,\cite{muta}. The
essential difference of the formula for the degree of divergence
(\ref{degree}) with the one related to QCD is the term $2n_{F}$. This term,
which supports the superficial convergence of the diagrams, precisely doubles
the one associated to QCD. This represents a drastic change in the behavior of
the divergences in the proposed model with respect to QCD or to simple NJL
models. After substituting the expression for the entering quantities $d_{G}$
can be rewritten as follows
\begin{align}
d_{G}  &  =\frac{1}{2}\sum_{i=1,...,N_{V}}n_{i}\left(  \frac{D-2}{2}%
(b_{i}+f_{i}+\delta_{i}-D\right)  -\nonumber\\
&  -\frac{(D-2)}{2}N_{B}-\frac{(D-2)}{2}N_{F}+D-\frac{N_{FP}}{2}.
\label{degree2}%
\end{align}
But, at $D=4$ for all the diagram in the model, it is valid $b_{i}%
+f_{i}+\delta_{i}=4-\delta_{i,4_{,}}.$ Thus, in this case the degree of
divergence can be written as
\begin{equation}
d_{G}=4-N_{G}-\frac{3}{2}N_{FP}-N_{F}-n_{gqq}. \label{degree3}%
\end{equation}
This result indicates that all the connected diagrams are superficially
convergent when the number of external lines is larger than 4. Therefore, the
model is power counting renormalizable in spite of including four fermion
interaction terms. Such a conclusion is surprising and thus the possibility of
completely construct the theory and investigate its predictions for the quark
masses must be carefully investigated. Consider now the set of numbers
$(N_{G},N_{FP},N_{F},n_{gqq})$ formed by the numbers of external gluon, ghost
and quark lines and also including in the fourth component the number of usual
gluon quark interaction Dirac vertices $n_{gqq} $. Employing the above formula
a finite set of superficially divergent Feynman diagrams can be determined for
the perturbative expansion.
%The allowed possibilities are enumerated in the table below.
%\begin{center}
%\begin{figure}[h]%
%\begin{tabular}
%[c]{|c||c||c||c|}\hline
%$(N_{G},N_{FP},N_{F},n_{gqq})$ & $d_{G}$ &   $(N_{G},N_{FP},N_{F},n_{gqq})$ &
%$d_{G}$\\\hline
%$\ \ \ \ (0,0,0,0)$ & $4$ &   $\ \ \ \ (0,0,2,,0),(0,0,2,1),(0,0,2,2)$ \,\,\,&
%$2,1,0$  \\\hline
%$\ \ \ \ (2,0,0,0)$ & $2$ &   $\ \ \ \ (0,0,4,0)$ & $0$\\\hline
%$\ \ \ \ (3,0,0,0)$ & $1$ &   $\ \ \ \ (1,0,2,0),(1,0,2,1)$ & $1,0$\\\hline
%$\ \ \ \ (4,0,0,0)$ & $0$ &   $\ \ \ \ (1,2,0)$ & $0$\\\hline
%$\ \ \ \ (0,2,0,0)$ & $1$ &   $\ \ \ \ (2,0,2)$ & $0$\\\hline
%\end{tabular}
%\caption{Table 1}%
%\label{table0}%
%\end{figure}
%\end{center}
At this point it should be underlined that precisely the new action terms
including two-gluons-two-quarks vertices, allowed the decreasing behavior of
the quark propagator at large momenta. This fact, permitted that most of the
Lagrangian terms defining the NJL models also became allowed to be included in
the Lagrangian without affecting the power counting renormalizability.
Henceforth, the mass generation properties embodied in usually
non-renormalizable NJL phenomenological theories, seem that can dynamically
work now in the considered context.

The sum of fourth order terms in the quark fields should be a general
expression being invariant under the symmetries of QCD, which became able to
allow the cancelation of the divergences appearing in each order of the
perturbative expansion. In the normal NJL theory, the usual Dirac propagator,
with its \textquotedblleft one over the momentum modulus\textquotedblright%
\ \ behavior at large momenta, makes the model non-renormalizable. It may be
also noted that the power counting rules for the model are similar to the ones
working in the simpler $\lambda\phi^{4}$ scalar field theory with the usual
scalar field propagator $\frac{1}{p^{2}}$ . This can be explicitly implemented
in the Lagrangian, simply by absorbing the six dimensional constants
$\varkappa_{q}$ in a redefinition of the quark fields. This change makes the
couplings associated to the NJL terms to becomes dimensionless constants. One
important remark following from the previous discussion can be added. Let us
assume that a renormalization procedure of pure massless QCD is reconsidered.
Then, the new added kind of vertices $\overline{\text{ }\Psi}$
$(x)\overleftarrow{D}$ $D\Psi(x)$, could be reasonably identified as possible
counterterms for this purpose, since they do not destroy power counting
renormalizability. This circumstance leads to the idea about the proposed
model could result to be physically equivalent to the quantized massless QCD.

\section{The renormalized Feynman expansion}

In this section we will develop the renormalized Feynman expansion of the
model. For this purpose the fields $A_{\mu}^{a},\chi^{\ast},\chi$,
$\overline{\psi}_{q}$ and $\psi_{q},q=1,2,...,6,$ in the bare action
(\ref{action}) will be substituted in terms of the renormalized fields through
the multiplicative mappings
\begin{align}
A_{\mu}^{a}  &  \rightarrow Z_{3}^{\frac{1}{2}}A_{\mu}^{a},\text{ \ }\\
\text{\ }\chi^{\ast}  &  \rightarrow\widetilde{Z}_{3}^{\frac{1}{2}}\chi^{\ast
},\text{ \ \ }\chi\rightarrow\widetilde{Z}_{3}^{\frac{1}{2}}\chi\text{ },\\
\overline{\psi}_{q}  &  \rightarrow Z_{q,2}^{\frac{1}{2}}\text{ }%
\overline{\psi}_{q}\text{ },\text{ \ \ \ \ }\psi_{q}\rightarrow Z_{q,2}%
^{\frac{1}{2}}\text{ }\psi_{q},\text{ \ \ \ \ }q=1,...,6,
\end{align}
and the bare coupling parameters of the action $g,$ $\alpha$ and
$\varkappa_{q}$ $(q=1,...,6),$ will be expressed in terms of their
renormalized couplings as follows%
\begin{align}
g-  &  >Z_{g}\text{ }g,\text{ \ \ }\\
\alpha-  &  >Z_{3}\alpha,\text{ \ \ }\\
\varkappa_{q}-  &  >Z_{\varkappa_{q}}\text{ }\varkappa_{q},\text{
\ \ \ \ }q=1,...,6,\\
\lambda_{f}-  &  >Z_{f}\text{ }\lambda_{f}\text{ \ \ \ \ \ \ \ }f=1,...,F.
\end{align}

With this substitutions done in the original action (\ref{action}), it can be
transformed in the sum of three terms as follows
\begin{align}
S  &  \mathcal{=}\text{ }S_{0}+S_{i}+S_{c}\nonumber\\
&  =\int dx^{D}\mathcal{(L}_{0}+\mathcal{L}_{i}+\mathcal{L}_{c}). \label{S}%
\end{align}
The quadratic in the fields term $S_{0}$, is the sum of the also quadratic
terms of the bare action, but in this case expressed in terms of the
renormalized fields and couplings
\begin{align}
S_{0}  &  =\sum_{n=1}^{n=3}S_{0}^{(n)}=\sum_{n=1}^{n=3}\int dx^{D}%
\mathcal{L}_{0}^{(n)},\\
S_{0}^{(1)}  &  =\int dx^{D}\frac{1}{2}A^{a\mu}(x)\delta^{ab}(g_{\mu\nu
}\partial^{2}-(1-\frac{1}{\alpha})\partial_{\mu}\partial_{\nu})A^{b\nu}(x),\\
S_{0}^{(2)}  &  =\int dx^{D}(\partial^{\mu}\chi^{a\ast}(x))\partial_{\mu}%
\chi^{a}(x),\\
S_{0}^{(3)}  &  =\sum_{q}\int dx^{D}\overline{\Psi}_{q}^{i}(x)\left(
\sigma\text{ }i\gamma^{\mu}\partial_{\mu}+\varkappa_{q}\partial^{2}\right)
\Psi_{q}^{j}(x).
\end{align}

The second term $S_{i}$ in (\ref{S}) is the sum of all Lagrangian terms of the
more than second order in the fields of the bare action, but in which the
fields are also substituted in terms of their renormalized values, and the
bare couplings are also substituted by the renormalized versions. Its
expression is
\begin{align}
S_{i}  &  =\sum_{n=1}^{n=7}S_{i}^{(n)}=\sum_{n=1}^{n=7}\int dx^{D}%
\mathcal{L}_{i}^{(n)},\\
S_{i}^{(1)}  &  =\int dx^{D}\frac{g}{2}f^{abc}(\partial_{\mu}A_{\nu}%
-\partial_{\nu}A_{\mu})A^{b\mu}A^{c\nu},\\
S_{i}^{(2)}  &  =-\int dx^{D}\frac{g^{2}}{4}f^{abe}f^{cde}A_{\mu}^{a}A_{\nu
}^{b}A^{c\mu}A^{d\nu},\\
S_{i}^{(3)}  &  =-\int dx^{D}gf^{abc}(\partial^{\mu}\chi^{a\ast})\chi
^{b}A_{\mu}^{c},\\
S_{i}^{(4)}  &  =-\sum_{q=1,...,6}\int dx^{D}\sigma g\overline{\Psi}_{q}%
^{i}\gamma^{\mu}T_{a}^{ij}\Psi_{q}^{j}A_{\mu}^{a},\\
S_{i}^{(5)}  &  =-i\sum_{q=1,...,6}\int dx^{D}\varkappa_{q}g\left(
\overline{\Psi}_{q}^{i}\partial^{\mu}(T_{a}^{ij}\Psi_{q}^{j}A_{\mu}%
^{a})+\overline{\Psi}_{q}^{i}A_{\mu}^{a}\partial^{\mu}(T_{a}^{ij}\Psi_{q}%
^{j})\right)  ,\\
S_{i}^{(6)}  &  =\sum_{q=1,...,6}\int dx^{D}\text{ }\varkappa_{f}\text{ }%
g^{2}\overline{\Psi}_{q}^{i}A^{c\mu}T_{c}^{ij}T_{b}^{jk}A_{\mu}^{b}\Psi
_{q}^{k},\\
S_{i}^{(7)}  &  =\sum_{f}\sum_{q_{1},q_{2},q_{_{3}},q_{_{4}}}\lambda_{f}\text{
}^{(f)}\Lambda_{(j_{2},r_{2},q_{_{2}})(j_{4},r_{4},q_{4})}^{(j_{1},r_{1}%
,q_{1})(j_{3,},r_{3},q_{3})}\times\nonumber\\
&  \int dx^{D}\text{ }\overline{\Psi}_{q_{_{1}}}^{j_{1},r_{1}}(x)\Psi
_{q_{_{2}}}^{j_{2},r_{2}}(x)\overline{\Psi}_{q_{_{3}}}^{j_{3},r_{3}}%
(x)\Psi_{q_{4}}^{j_{4},r_{4}}\text{ }(x),
\end{align}
Finally, the third term $S_{c}$ is the difference between the renormalized
action and the bare action expressed in terms of the renormalized fields and
couplings. It can be written in the form

\begin{center}%
\begin{align}
S_{c}  &  =\sum_{n=1}^{n=11}S_{c}^{(n)}=\sum_{n=1}^{n=11}\int dx^{D}%
\mathcal{L}_{c}^{(n)},\\
S_{c}^{(1)}  &  =(Z_{3}-1)\int dx^{D}\frac{1}{2}A^{a\mu}\delta^{ab}(g_{\mu\nu
}\partial^{2}-\partial_{\mu}\partial_{\nu})A^{b\nu},\\
S_{c}^{(2)}  &  =(\widetilde{Z}_{3}-1)\int dx^{D}(\partial^{\mu}\chi^{a\ast
})\partial_{\mu}\chi^{a},\\
S_{c}^{(3)}  &  =\sigma\sum_{q}(Z_{2}^{q}-1)\int dx^{D}\overline{\Psi}_{q}%
^{i}(x)\text{ }i\gamma^{\mu}\partial_{\mu}\Psi_{q}^{i}(x),\\
S_{c}^{(4)}  &  =(Z_{1}-1)\int dx^{D}\text{ }\frac{g}{2}\text{ }%
f^{abc}(\partial_{\mu}A_{\nu}-\partial_{\nu}A_{\mu})A^{b\mu}A^{c\nu},\\
S_{c}^{(5)}  &  =-(Z_{4}-1)\int dx^{D}\text{ }\frac{g^{2}}{4}\text{ }%
f^{abe}f^{cde}A_{\mu}^{a}A_{\nu}^{b}A^{c\mu}A^{d\nu},
\end{align}

\begin{align}
S_{c}^{(6)} &  =-(\widetilde{Z}_{1}-1)\int dx^{D}g\text{ }f^{abc}%
(\partial^{\mu}\chi^{a\ast})\chi^{b}A_{\mu}^{c},\\
S_{c}^{(7)} &  =-\sum_{q=1,...,6}(Z_{1F}^{q}-1)\int dx^{D}\sigma g\text{
}\overline{\Psi}_{q}^{i}\gamma^{\mu}T_{a}^{ij}\Psi_{q}^{j}A_{\mu}^{a},\\
S_{c}^{(8)} &  =\sum_{q=1,...,6}(Z_{\varkappa_{q}}^{1}-1)\int dx^{D}\text{
}\varkappa_{q}\text{ }\overline{\Psi}_{q}^{i}\partial^{2}\Psi_{q}^{j},\\
S_{c}^{(9)} &  =-i\sum_{q=1,...,6}(Z_{\varkappa_{q}}^{2}-1)\int dx^{D}\text{
}\varkappa_{q}g\text{ }\left(  \overline{\Psi}_{q}^{i}\partial^{\mu}%
(T_{a}^{ij}\Psi_{q}^{j}A_{\mu}^{a})+\overline{\Psi}_{q}^{i}A_{\mu}^{a}%
\partial^{\mu}(T_{a}^{ij}\Psi_{q}^{j})\right)  ,\\
S_{c}^{(10)} &  =\sum_{q=1,...,6}(Z_{\varkappa_{q}}^{3}-1)\int dx^{D}\text{
}\varkappa_{f}\text{ }g^{2}\overline{\Psi}_{q}^{i}A^{c\mu}T_{c}^{ij}T_{b}%
^{jk}A_{\mu}^{b}\Psi_{q}^{k},\\
S_{c}^{(11)} &  =\sum_{f}\sum_{q_{1},q_{2},q_{_{3}},q_{_{4}}}(Z_{\lambda
}(Z_{2,q_{1}}Z_{2,q_{2}}Z_{2,q_{3}}Z_{2,q_{4}})^{^{\frac{1}{2}}}%
-1)\frac{\lambda_{f}}{4}\text{ }^{(f)}\Lambda_{(j_{2},r_{2},q_{_{2}}%
)(j_{4},r_{4},q_{4})}^{(j_{1},r_{1},q_{1})(j_{3,},r_{3},q_{3})}\times
\nonumber\\
&  \int dx^{D}\text{ }\overline{\Psi}_{q_{_{1}}}^{j_{1},r_{1}}(x)\Psi
_{q_{_{2}}}^{j_{2},r_{2}}(x)\overline{\Psi}_{q_{_{3}}}^{j_{3},r_{3}}%
(x)\Psi_{q_{4}}^{j_{4},r_{4}}\text{ }(x),
\end{align}

\end{center}

in which the following definitions of the appearing new parameters had been
considered
\begin{align}
Z_{1}  &  =Z_{g}Z_{3}^{\frac{3}{2}},\\
\widetilde{Z}_{1}  &  =Z_{g}\widetilde{Z}_{3}Z_{3}^{\frac{1}{2}},\\
Z_{4}  &  =Z_{g}^{2}Z_{3}^{2},\\
Z_{1F}^{q}  &  =Z_{g}Z_{2}^{q}Z_{3}^{\frac{1}{2}},\text{ \ \ \ \ \ \ \ }%
q=1,...,6,\\
Z_{\varkappa_{q}}^{1}  &  =Z_{\varkappa_{q}}Z_{q,2},\text{
\ \ \ \ \ \ \ \ \ \ }q=1,...,6,\\
Z_{\varkappa_{q}}^{2}  &  =Z_{\varkappa_{q}}Z_{q,2}Z_{3}^{\frac{1}{2}}%
Z_{g},\text{ \ \ }q=1,...,6,\\
Z_{\varkappa_{q}}^{3}  &  =Z_{\varkappa_{q}}Z_{q,2}Z_{3}Z_{g}^{2},\text{
\ \ \ }q=1,...,6.
\end{align}

\subsection{The generating functional for the Feynman diagrams}

Henceforth, considering the above definitions for the action of the model and
introducing the auxiliary external sources for the three fields, the explicit
form of the generating functional of the perturbative expansion can be written
in the form
\begin{align}
Z[j,\eta,\overline{\eta},\xi,\xi^{\ast}]  &  =\frac{1}{\mathcal{N}}\int
D[A,\overline{\Psi}_{q},\Psi_{q},{\small {\large \chi}}^{\ast}%
,{\small {\large \chi}}]\exp[i\text{{}}{\LARGE [}S[A,\Psi_{q},\overline{\Psi
}_{q},{\small {\large \chi}}^{\ast},{\small {\large \chi}}]+\nonumber\\
&  +\int dx^{D}{\Large (}{\small {\large j(x)A(x)+\xi}}^{\ast}%
{\small {\large (x)\chi(x)+\chi}}^{\ast}{\small {\large (x)\xi(x)}%
+}\nonumber\\
&  \sum_{q=1,...,6}{\large (}\overline{{\small \eta}}_{q}(x)\Psi
_{q}(x)+\overline{\Psi}_{q}(x)\eta_{q}(x){\small {\large )}}{\Large )}%
{\huge ]}\nonumber\\
&  =\frac{1}{\mathcal{N}}\exp{\Huge [}i\text{ }{\huge (}S{\small [}%
\frac{\delta}{i\text{ }\delta j}{\small ,\frac{\delta}{-i\text{ }\delta
\eta_{q}},}\frac{\delta}{i\text{ }\delta\overline{\eta}_{q}},{\small \frac
{\delta}{-i\text{ }\delta\xi}},\frac{\delta}{i\text{ }\delta\xi_{q}^{\ast}%
}{\small ]-}\nonumber\\
&  \text{ \ \ \ \ \ \ \ \ \ \ \ \ \ \ \ \ \ \ }-{\small S}_{0}{\small [\frac
{\delta}{i\text{ }\delta j},\frac{\delta}{-i\text{ }\delta\eta_{q}}%
,\frac{\delta}{i\text{ }\delta\overline{\eta}_{q}},\frac{\delta}{-i\text{
}\delta\xi},\frac{\delta}{i\text{ }\delta\xi_{q}^{\ast}}]}{\huge )}%
{\Huge ]}\times\nonumber\\
&  \text{ \ \ \ \ \ \ \ \ }\times\text{\ }Z^{(0)}[j,\eta,\overline{\eta}%
,\xi,\xi^{\ast}].
\end{align}
where $\mathcal{N}$ is chosen for assuring that $Z$ vanish when all the
sources are equal to zero. In the above expression $Z^{(0)}$ is the free
partition function, which is the product of eight separate factors for gluons,
ghosts and for each type of quark, in the form%
\begin{align}
Z^{(0)}[j,\eta,\overline{\eta},\xi,\xi^{\ast}]  &  =Z_{G}^{(0)}[j]\times
Z_{GH}^{(0)}[\xi,\xi^{\ast}]\times%
%TCIMACRO{\dprod \limits_{q=1,...,6}}%
%BeginExpansion
{\displaystyle\prod\limits_{q=1,...,6}}
%EndExpansion
Z_{q}^{(0)}[\eta_{q},\overline{\eta}_{q}].\nonumber\\
&  =\int D[A,\overline{\Psi}_{q},\Psi_{q},{\small {\large \chi}}^{\ast
},{\small {\large \chi}}]\exp{\Large [}i\text{ }{\LARGE (}S_{0}[A,\Psi
_{q},\overline{\Psi}_{q},{\small {\large \chi}}^{\ast},{\small {\large \chi}%
}]+\nonumber\\
&  +{\small {\large \int dx^{D}{\Large (}}}j(x)A(x)+{\small {\large \xi}%
}^{\ast}{\small {\large (x)\chi(x)+\chi}}^{\ast}{\small {\large (x)\xi(x)}%
+}\nonumber\\
&  \sum_{q=1,...,6}{\large (}\overline{{\small \eta}}_{q}(x)\Psi
_{q}(x)+\overline{\Psi}_{q}(x)\eta_{q}(x){\small {\large )}}{\Large )]}%
,\nonumber\\
&  =\int D[A]\exp{\Huge [}i{\huge (}\int dx^{D}\frac{1}{2}A^{a\mu}%
(x)\delta^{ab}(g_{\mu\nu}\partial^{2}-(1-\frac{1}{\alpha})\partial_{\mu
}\partial_{\nu})A^{b\nu}(x)+\nonumber\\
&  \text{ \ \ \ \ \ \ \ \ \ \ \ \ \ \ \ \ \ \ \ \ \ \ \ \ \ \ }\int
dx^{D}\text{ }{\small {\large j}}_{\mu}(x){\small {\large A}}^{\mu
}(x){\huge )}{\Huge ]}\times\nonumber\\
&  \int D[{\small {\large \chi}}^{\ast},{\small {\large \chi}}]\exp
{\huge [}i{\LARGE (}\int dx^{D}(\partial^{\mu}\chi^{a\ast}(x))\partial_{\mu
}\chi^{a}(x)+\nonumber\\
&  \text{ \ \ \ \ \ \ \ \ \ \ \ \ \ \ \ \ \ \ \ \ \ \ \ \ \ \ \ }+\int
dx^{D}\text{ }\left(  {\small \xi}^{a\ast}{\small (x)\chi}^{a}{\small (x)+\chi
}^{a\ast}{\small (x)\xi}^{a}{\small (x)}\right)  {\LARGE )}{\huge ]}%
\times\nonumber\\
&
%TCIMACRO{\dprod \limits_{q}}%
%BeginExpansion
{\displaystyle\prod\limits_{q}}
%EndExpansion
\int D[\overline{\Psi}_{q},\Psi_{q}]\exp{\Huge [}i\text{ }{\LARGE (}\int
dx^{D}\overline{\Psi}_{q}^{i}(x)\left(  \sigma\text{ }i\gamma^{\mu}%
\partial_{\mu}+\varkappa_{q}\partial^{2}\right)  \Psi_{q}^{j}(x).{\small +}%
\nonumber\\
&  \text{ \ \ \ \ \ \ \ \ \ \ \ \ \ \ \ \ \ \ \ \ \ \ \ \ \ \ \ \ \ \ \ \ \ }%
\times\int dx^{D}\left(  \overline{{\small \eta}}_{q}(x)\Psi_{q}%
(x)+\overline{\Psi}_{q}(x)\eta_{q}(x)\right)  {\LARGE )}{\Huge ].}%
\end{align}
The expressions for the gluon, ghost and quark free partition functions follow
in the forms
\begin{align}
Z_{G}^{(0)}[j]  &  =\exp\left(  \frac{i}{2}\int dx^{D}dy^{D}\text{ }j^{a\mu
}(x)D_{\mu\nu}^{ab}(x-y)j^{b\nu}(x)\right)  ,\\
D_{\mu\nu}^{ab}(x.y)  &  =\delta^{ab}\int\frac{dk^{D}}{(2\pi)^{D}}\frac
{\exp(-ik(x-y))}{k^{2}+i\epsilon}(g_{\mu\nu}-(1-\alpha)k_{\mu}k_{\nu}),\\
Z_{GH}^{(0)}[\xi,\xi^{\ast}]  &  =\exp\left(  \frac{i}{2}\int dx^{D}%
dy^{D}\text{ }\xi^{\ast a}(x)D^{ab}(x-y)\xi^{b}(x)\right)  ,\\
D^{ab}(x.y)  &  =-\delta^{ab}\int\frac{dk^{D}}{(2\pi)^{D}}\frac{\exp
(-ik(x-y))}{k^{2}+i\epsilon},\\
Z_{q}^{(0)}[\eta_{q},\overline{\eta}_{q}]  &  =\exp\left(  i\int dx^{D}%
dy^{D}\text{ }\overline{\eta}_{q}^{r_{1}}(x)S_{q}^{r_{1}r_{2}}(x-y)\eta
_{q}^{r_{2}}(x)\right)  ,\text{ \ \ }q=1,...,6,\\
S_{q}^{r_{1}r_{2}}(x-y)  &  =\int\frac{dk^{D}}{(2\pi)^{D}}\frac{-\sigma
}{\gamma_{\mu}k^{\mu}-\sigma\varkappa_{q}k^{2}}\exp(-ik(x-y)).
\end{align}

\subsection{The Feynman diagram expansion}

\begin{figure}[h]
\includegraphics[scale=0.55]{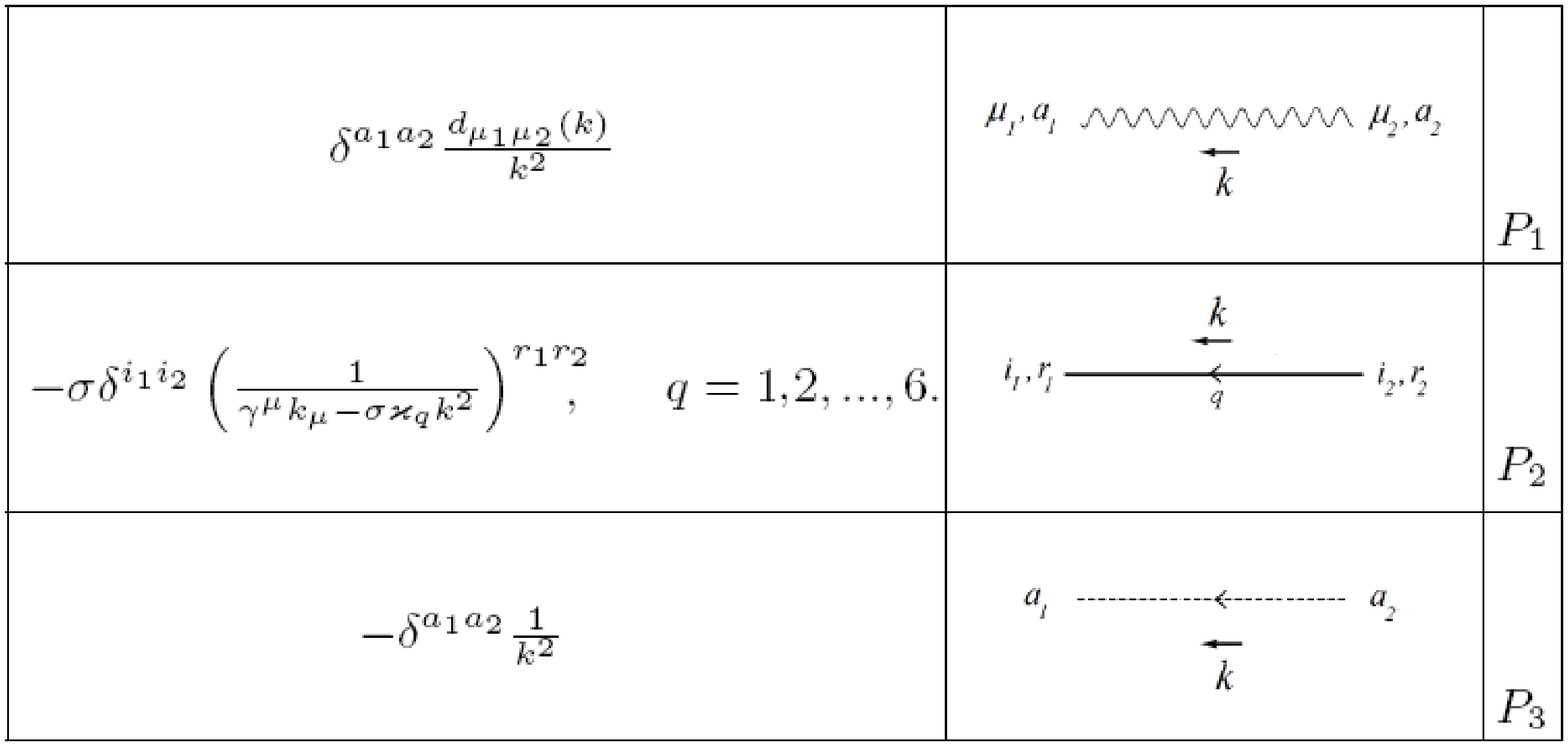}\caption{The gluon, ghost and quark
propagators for the renormalized fields.}%
\label{table1}%
\end{figure}

The just defined generating functional leads to the Feynman rules which are
presented in detail in Appendix A. The free propagators for the gluon, ghost
and quark fields, in these rules and their corresponding diagrams are shown in
figure \ref{table1}. The essential difference with QCD propagators is in the form of
\ the six quark ones, which now decrease at large momenta in a quadratic form.
The Feynman rules for the seven vertices of cubic and quartic order in the
fields coming from the bare action, after expressed in terms of the
renormalized fields and couplings are depicted in the figure 4 of the Appendix
A. The three first vertices $V_{1}$ to $V_{3}$ are identical to the ones in
QCD, and the fourth one $V_{4}$ only includes as a modification the allowed
change $\sigma$ in the sign of the Dirac action. The vertices $V_{5}$ and
$V_{6}$ are the ones coming from the new local, gauge invariant and
renormalizable terms added to action for each $flavour$ value $q$. The
diagrams and their corresponding analytic expressions for the counterterm
vertices are presented in the figure 5 of the Appendix A. The standard
expressions $V$ and $W$ for the vertices of QCD and the function $d_{\mu\nu
}(k)$ defining the gluon propagator appearing in the above graphics for
propagators, vertices and counterterms have the explicit forms \cite{muta}%
\begin{align}
V_{\mu_{1}\mu_{2}\mu_{3}}^{a_{1}a_{2}a_{3}}(k_{1},k_{2},k_{3})  &  =\text{
}f^{a_{1}a_{2}a_{3}}{\Large (}(k_{1}-k_{2})_{\mu_{3}}g_{\mu_{1}\mu_{2}%
})+(k_{2}-k_{3})_{\mu_{1}}g_{\mu_{2}\mu_{3}})+(k_{3}-k_{1})_{\mu_{2}}%
g_{\mu_{1}\mu_{3}}{\Large ),}\\
W_{\mu_{1}\mu_{2}\mu_{3}\mu_{4}}^{a_{1}a_{2}a_{3}a_{4}}  &  =(f^{13,24}%
-f^{14,32})g_{\mu_{1}\mu_{2}}g_{\mu_{3}\mu_{4}}+(f^{12,34}-f^{14,23}%
)g_{\mu_{1}\mu_{3}}g_{\mu_{2}\mu_{4}}+\nonumber\\
&  \text{ \ \ \ }(f^{13,42}-f^{12,34})g_{\mu_{1}\mu_{4}}g_{\mu_{3}\mu_{2}},\\
f^{\text{ }ij,kl}  &  =f^{\text{ }a_{i}a_{j}a}f^{\text{ }a_{k}a_{l}a},\\
d_{\mu\nu}(k)  &  =(g_{\mu\nu}-(1-\alpha)\frac{k_{\mu}k_{\nu}}{k^{2}}).
\end{align}

\section{Primitively divergent graphs and their divergence}

\begin{figure}[h]
\includegraphics[scale=0.25]{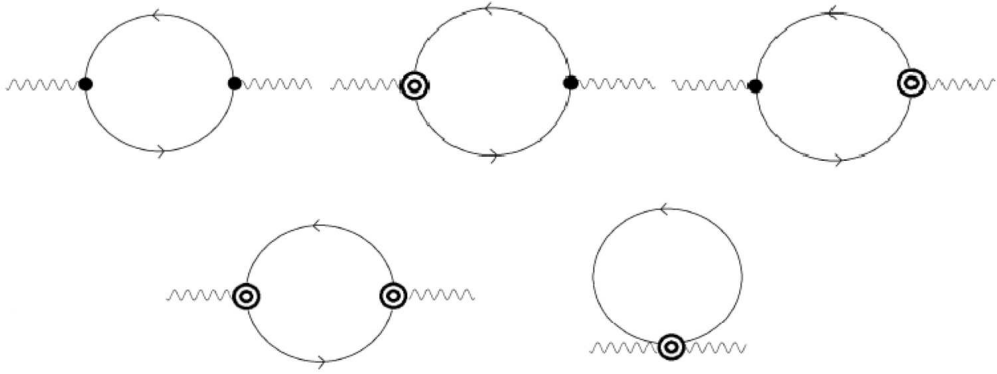}\caption{The diagrams with two
gluon legs: $(2,0,0,2),2\times(2,0,0,1),2\times(2,0,0,0)$, contributing to the
primitive divergences. Only diagrams not already present in gluodynamics are
shown. }%
\label{gluonself}%
\end{figure}

Let us consider in this section the primitively divergent graphs of the model.
We will evaluate those ones corresponding to two external gluon legs. \ That
is, being associated to the sets of four numbers $(N_{G},N_{FP},N_{F}%
,n_{gqq})$ taking the specific values $(2,0,0,2),2\times(2,0,0,1)$ and
$2\times(2,0,0,0)$. Note that there are sets of four numbers corresponding to
two graphs. This is the meaning of the factors "2" of the sets $(2,0,0,1)$ and
$(2,0,0,0)$. They are depicted in the figure \ref{gluonself}\ below. The
divergences of the rest of primitively divergent graphs, due to their large
number will be considered in a separate work. After evaluated the whole set of
divergent counterterms in the one loop approximation, in addition to the ones
present in pure gluodynamics (See \cite{muta}), this information will allow to
study the renormalization group properties of the model in the one loop
approximation. However, the evaluation will started in this work by
calculating the one loop divergence associated to the gluon self-energy
diagrams shown in figure \ref{gluonself} . After writing the analytical expressions by using the Feynman
rules the set of five terms can be written in the form
\begin{align}
\Pi_{\mu\nu}^{(1)}(k) &  =(-\frac{\delta^{ab}g^{2}}{2i})\sum_{q}\int\frac
{dp}{(2\pi)^{D}}T_{r}^{D}[\gamma_{\mu}S_{m_{q}}(k+p)\gamma_{\nu}S_{m_{q}%
}(p)+\gamma_{\mu}S_{0}(k+p)\gamma_{\nu}S_{0}(p)-\nonumber\\
&  \text{ \ \ \ \ \ \ \ \ \ \ \ \ \ \ \ \ \ \ \ \ }-\gamma_{\mu}S_{m_{q}%
}(k+p)\gamma_{\nu}S_{0}(p)-\gamma_{\mu}S_{0}(k+p)\gamma_{\nu}S_{m_{q}}(p)],\\
\Pi_{\mu\nu}^{(2)}(k) &  =(-\frac{\delta^{ab}g^{2}}{2i})\sum_{q}%
\sigma\varkappa_{q}\int\frac{dp}{(2\pi)^{D}}T_{r}^{D}[S_{m_{q}}(k+p)\gamma
_{\nu}S_{m_{q}}(p)+S_{0}(k+p)\gamma_{\nu}S_{0}(p)-\nonumber\\
&  \text{ \ \ \ \ \ \ \ \ \ \ \ \ \ \ \ \ \ \ \ \ }-S_{m_{q}}(k+p)\gamma_{\nu
}S_{0}(p)-S_{0}(k+p)\gamma_{\nu}S_{m_{q}}(p)](2p_{\mu}+k_{\mu}),\\
\Pi_{\mu\nu}^{(3)}(k) &  =(-\frac{\delta^{ab}g^{2}}{2i})\sum_{q}%
\sigma\varkappa_{q}\int\frac{dp}{(2\pi)^{D}}T_{r}^{D}[S_{m_{q}}(k+p)\gamma
_{\mu}S_{m_{q}}(p)+S_{0}(k+p)\gamma_{\mu}S_{0}(p)-\nonumber\\
&  \text{ \ \ \ \ \ \ \ \ \ \ \ \ \ \ \ \ \ \ \ \ \ \ \ }-S_{_{m}}%
(k+p)\gamma_{\mu}S_{0}(p)-S_{0}(k+p)\gamma_{\mu}S_{_{m}}(p)](2p_{\nu}+k_{\nu
}),\\
\Pi_{\mu\nu}^{(4)}(k) &  =(-\frac{\delta^{ab}g^{2}}{2i})\sum_{q}%
(\sigma\varkappa_{q})^{2}\int\frac{dp}{(2\pi)^{D}}T_{r}^{D}[S_{m_{q}%
}(k+p)S_{m_{q}}(p)+S_{0}(k+p)S_{0}(p)-\nonumber\\
&  \text{ \ \ \ \ \ \ \ \ \ \ \ \ \ \ \ \ \ \ \ \ \ \ \ \ }-S_{m_{q}%
}(k+p)S_{0}(p)-S_{0}(k+p)S_{m_{q}}(p)](2p_{\mu}+k_{\mu})(2p_{\nu}+k_{\nu}),\\
\Pi_{\mu\nu}^{(5)}(k) &  =(-\frac{\delta^{ab}g^{2}}{i})\sum_{q}(\sigma
\varkappa_{q})^{2}\int\frac{dp}{(2\pi)^{D}}T_{r}^{D}[S_{_{m}}(p)-S_{0}(p)],
\end{align}
where $m_{q}=\frac{\sigma}{\varkappa_{q}},q=1,...,6$, are the quark masses and
the $S_{m_{q}}(p)$ and $S_{0}(p)$ are Dirac free propagators
\begin{align}
S_{m_{p}}(p) &  =\frac{1}{m_{p}-p_{\alpha}\gamma^{\alpha}},\\
S_{0}(p) &  =\frac{1}{-p_{\alpha}\gamma^{\alpha}}.
\end{align}

The above quantities can be evaluated from the general integrals%
\begin{align}
I_{\mu\nu}^{(1)}(k,m_{1},m_{2}) &  =\int\frac{dp}{(2\pi)^{D}}T_{r}^{D}%
[\gamma_{\mu}S_{m_{1}}(k+p)\gamma_{\nu}S_{m_{2}}(p)]\nonumber\\
&  =\int\frac{dp}{(2\pi)^{D}}4\frac{(m_{1}m_{2}-(p+k).kg_{\mu\nu}+2p_{\mu\nu
}+p_{\mu}k_{\nu}+k_{\mu}p_{\nu})}{(m_{1}^{2}-(p+k)^{2})(m_{2}^{2}-(p)^{2})},\\
I_{\mu\nu}^{(2)}(k,m_{1},m_{2}) &  =-\int\frac{dp}{(2\pi)^{D}}T_{r}%
^{D}[S_{m_{1}}(k+p)\gamma_{\nu}S_{m_{2}}(p)](2p_{\mu}+k_{\mu})\nonumber\\
&  =-\int\frac{dp}{(2\pi)^{D}}4\frac{(m_{2}k_{\mu}k_{\nu}+2(m_{1}%
+m_{2}).p_{\mu}p_{\nu}+(m_{1}+m_{2})p_{\nu}k_{\mu}+2m_{2}p_{\mu}k_{\nu}%
)}{(m_{1}^{2}-(p+k)^{2})(m_{2}^{2}-(p)^{2})},\\
I_{\mu\nu}^{(3)}(k,m_{1},m_{2}) &  =-\int\frac{dp}{(2\pi)^{D}}T_{r}%
^{D}[S_{m_{1}}(k+p)\gamma_{\mu}S_{m_{2}}(p)](2p_{\nu}+k_{\nu})\nonumber\\
&  =-\int\frac{dp}{(2\pi)^{D}}4\frac{(m_{2}k_{\mu}k_{\nu}+2(m_{1}%
+m_{2}).p_{\mu}p_{\nu}+(m_{1}+m_{2})p_{\mu}k_{\nu}+2m_{2}p_{\nu}k_{\mu}%
)}{(m_{1}^{2}-(p+k)^{2})(m_{2}^{2}-(p)^{2})},\\
I_{\mu\nu}^{(4)}(k,m_{1},m_{2}) &  =\int\frac{dp}{(2\pi)^{D}}T_{r}%
^{D}[S_{m_{1}}(k+p)S_{m_{2}}(p)](2p_{\mu}+k_{\mu})(2p_{\nu}+k_{\nu
})\nonumber\\
&  =\int\frac{dp}{(2\pi)^{D}}4\frac{(m_{1}m_{2}+p^{2}+p.k)(2p_{\mu}+k_{\mu
})(2p_{\nu}+k_{\nu})}{(m_{1}^{2}-(p+k)^{2})(m_{2}^{2}-(p)^{2})},\\
I_{\mu\nu}^{(5)}(k,m) &  =\int\frac{dp}{(2\pi)^{D}}T_{r}^{D}[S_{_{m}}%
(p)-S_{0}(p)]\nonumber\\
&  =\sigma\text{ }m\int\frac{dp}{(2\pi)^{D}}4\frac{1}{(p^{2}-m^{2})},
\end{align}
which are functions of two general values of the masses $m_{1}$ and $m_{2}$ .
Only their values at the mass pairs $(m_{1},m_{2})=(m_{q}$,$m_{q}),(m_{q}%
$,$0),(0$,$m_{q}),(0$,$0)$ are needed. The appearing momentum integrals can be
transformed by using the Feynman trick
\begin{align}
\frac{1}{(m_{1}^{2}-(p+k)^{2})(m_{2}^{2}-(p)^{2})} &  =\int_{0}^{1}dx\frac
{1}{[(p+x\text{ }k)^{2}-R_{m_{1}m_{2}}^{2}(x)]^{2}},\\
R_{m_{1}m_{2}}^{2}(x) &  =k^{2}x(x-1)+m_{2}^{2}(1-x)+m_{1}^{2}-i\epsilon.
\end{align}
After substituting the above representation for the denominators in the
defined four integrals $I$ , making the integration variable shift
$p\rightarrow p+x$ $k$ and using the momentum integrals \cite{narison}
\begin{align}
\int\frac{dp}{(2\pi)^{D}}\frac{1}{[p^{2}-R_{m_{1}m_{2}}^{2}(x)^{2}]^{2}} &
=\frac{i}{(16\pi^{2})^{\frac{D}{4}}}\frac{\Gamma(2-\frac{D}{2})}{\Gamma
(2)}(R_{m_{1}m_{2}}^{2}(x))^{\frac{D}{2}-2},\\
\int\frac{dp}{(2\pi)^{D}}\frac{p_{\mu}p_{\nu}}{[p^{2}-R_{m_{1}m_{2}}%
^{2}(x)]^{2}} &  =\frac{iR_{m_{1}m_{2}}^{2}(x)}{(16\pi^{2})^{\frac{D}{4}}%
}\frac{\Gamma(2-\frac{D}{2})}{D\Gamma(2)}R_{m_{1}m_{2}}^{2}(x)^{\frac{D}{2}%
-2},\\
\int\frac{dp}{(2\pi)^{D}}\frac{p^{2}p_{\mu}p_{\nu}}{[p^{2}-R_{m_{1}m_{2}}%
^{2}(x)]^{2}} &  =\frac{1}{D}g_{\mu\nu}\int\frac{dp}{(2\pi)^{D}}\frac
{(p^{2})^{2}}{[p^{2}-R_{m_{1}m_{2}}^{2}(x)^{2}]^{2}}\nonumber\\
&  =g_{\mu\nu}\frac{i(D+2)R_{m_{1}m_{2}}^{2}(x)}{(D-2)(16\pi^{2})^{\frac{D}%
{4}}}\frac{\Gamma(2-\frac{D}{2})}{\Gamma(2)}R_{m_{1}m_{2}}^{2}(x)^{\frac{D}%
{2}-2},
\end{align}
the $I$ integrals can be rewritten as
\begin{align}
I_{\mu\nu}^{(1)}(k,m_{1},m_{2}) &  =\frac{i\text{ }\Gamma(2-\frac{D}{2}%
)}{(16\pi^{2})^{\frac{D}{4}}\Gamma(2)}\int_{0}^{1}dx\text{ }(R_{m_{1}m_{2}%
}^{2}(x))^{\frac{D}{2}-2}\text{ }4{\large [}-R_{m_{1}m_{2}}^{2}(x)g_{\mu\nu
}+\nonumber\\
&  \text{ \ \ \ \ \ \ \ \ \ \ \ \ \ \ \ \ \ \ \ \ \ \ \ \ \ \ \ \ \ }%
+m_{1}m_{2}+x(1-x)k^{2})+2x(x-1)k_{\mu}k_{\nu}{\large ]},\\
I_{\mu\nu}^{(2)}(k,m_{1},m_{2}) &  =\frac{i\text{ }\Gamma(2-\frac{D}{2}%
)}{(16\pi^{2})^{\frac{D}{4}}\Gamma(2)}\int_{0}^{1}dx\text{ }(R_{m_{1}m_{2}%
}^{2}(x))^{\frac{D}{2}-2}\text{ }4{\large [}g_{\mu\nu}\frac{2}{D-2}%
(m_{1}+m_{2})\times\nonumber\\
&  \text{ \ \ \ \ \ \ \ \ \ \ \ \ \ \ \ \ \ \ \ \ \ \ \ \ \ \ \ \ \ }%
\times(-x(x-1)k^{2}+m_{2}^{2}(1-x)+m_{1}^{2}x-i\epsilon)+\\
&  \text{ \ \ \ \ \ \ \ \ \ \ \ \ \ \ \ \ \ \ \ \ \ \ \ \ \ \ }+k_{\mu}k_{\nu
}\ (m_{2}(1-3x)-m_{1}x+2(m_{1}+m_{2})\ x^{2}{\large ]},\nonumber\\
I_{\mu\nu}^{(3)}(k,m_{1},m_{2}) &  =\frac{i\text{ }\Gamma(2-\frac{D}{2}%
)}{(16\pi^{2})^{\frac{D}{4}}\Gamma(2)}\int_{0}^{1}dx\text{ }(R_{m_{1}m_{2}%
}^{2}(x))^{\frac{D}{2}-2}\text{ }4{\large [}g_{\mu\nu}\frac{2}{D-2}%
(m_{1}+m_{2})\times\nonumber\\
&  \text{ \ \ \ \ \ \ \ \ \ \ \ \ \ \ \ \ \ \ \ \ \ \ \ \ \ \ \ \ \ \ \ \ }%
\times(-x(x-1)k^{2}+m_{2}^{2}(1-x)+m_{1}^{2}x-i\epsilon)+\nonumber\\
&  \text{ \ \ \ \ \ \ \ \ \ \ \ \ \ \ \ \ \ \ \ \ \ \ \ \ \ \ \ \ }+k_{\mu
}k_{\nu}\ (m_{2}(1-3x)-m_{1}x+2(m_{1}+m_{2})\ x^{2}{\large ]},\\
I_{\mu\nu}^{(4)}(k,m_{1},m_{2}) &  =\frac{i\text{ }\Gamma(2-\frac{D}{2}%
)}{(16\pi^{2})^{\frac{D}{4}}\Gamma(2)}\int_{0}^{1}dx\text{ }(R_{m_{1}m_{2}%
}^{2}(x))^{\frac{D}{2}-2}\text{ }4{\LARGE [}g_{\mu\nu}{\Large (}\frac
{4(D-2)}{D(D-2)}\text{ }(R^{2}(x))^{2}+\nonumber\\
&  \text{
\ \ \ \ \ \ \ \ \ \ \ \ \ \ \ \ \ \ \ \ \ \ \ \ \ \ \ \ \ \ \ \ \ \ \ \ \ }%
+4\frac{(m_{1}m_{2}+x(x-1)k^{2})}{D-2}R_{m_{1}m_{2}}^{2}(x){\LARGE )}\ +\\
&  \text{ \ \ \ \ \ \ \ \ \ \ }+k_{\mu}k_{\nu}(1-2x)^{2}{\Large (}\frac
{(D+4)}{(D-2)}R_{m_{1}m_{2}}^{2}(x)+x(x-1)k^{2}+m_{1}m_{2}{\LARGE )}%
{\Large ],}\nonumber\\
I_{\mu\nu}^{(5)}(k,m) &  =-\sigma\frac{i}{(16\pi^{2})^{\frac{D}{4}}}%
m^{\frac{D}{2}-1}\frac{\Gamma(2-\frac{D}{2})}{(1-\frac{D}{2})\Gamma(2)}.
\end{align}
The obtained expressions show their divergence determined by the
$D\rightarrow4$ limit of the factor
\begin{equation}
\Gamma(2-\frac{D}{2})=\Gamma(-\frac{\epsilon}{2})=-\frac{2}{\epsilon}%
(1+\frac{\gamma}{2}\epsilon+O(\epsilon^{2}),
\end{equation}
with the definition $D=4-\epsilon.$ Thus, the divergent part of the
self-energy contributions under consideration can be expressed as:%
\begin{align}
\Pi_{\mu\nu}^{(1\operatorname{div})}(k) &  =P[\sum_{q}(-\frac{\delta^{ab}%
g^{2}}{2i})(I_{\mu\nu}^{(1)}(k,m_{q},m_{q})+I_{\mu\nu}^{(1)}(k,m_{q}%
,0)-\nonumber\\
&  \text{ \ \ \ }-I_{\mu\nu}^{(1)}(k,0,m_{q})-I_{\mu\nu}^{(1)}%
(k,0,0))]\nonumber\\
&  =\frac{1}{16\pi^{2}}\frac{2}{\epsilon}(\frac{\delta^{ab}g^{2}}{2})\sum
_{q}4\times(-m_{q}^{2})\text{ }g_{\mu\nu},\\
\Pi_{\mu\nu}^{(2\operatorname{div})}(k) &  =-P[\sum_{q}(-\frac{\delta
^{ab}g^{2}}{2})\sigma\varkappa_{q}(I_{\mu\nu}^{(2)}(k,m_{q},m_{q})+I_{\mu\nu
}^{(2)}(k,m_{q},0)-\nonumber\\
&  \text{ \ \ }-I_{\mu\nu}^{(2)}(k,0,m_{q})-I_{\mu\nu}^{(2)}%
(k,0,0))]\nonumber\\
&  =\frac{1}{16\pi^{2}}\frac{2}{\epsilon}\frac{\delta^{ab}g^{2}}{2}\sum
_{q}4\times(-m_{q}^{2})\text{ }g_{\mu\nu},\\
\Pi_{\mu\nu}^{(3\operatorname{div})}(k) &  =-P[\sum_{q}(-\frac{\delta
^{ab}g^{2}}{2i})\sigma\varkappa_{q}(I_{\mu\nu}^{(3)}(k,m_{q},m_{q})+I_{\mu\nu
}^{(3)}(k,m_{q},0)-\nonumber\\
&  \text{ \ \ \ }-I_{\mu\nu}^{(3)}(k,0,m_{q})-I_{\mu\nu}^{(3)}%
(k,0,0))]\nonumber\\
&  =\frac{1}{16\pi^{2}}\frac{2}{\epsilon}\frac{\delta^{ab}g^{2}}{2}\sum
_{q}4\times(-m_{q}^{2})\text{ }g_{\mu\nu},\\
\Pi_{\mu\nu}^{(4\operatorname{div})}(k) &  =P[\sum_{q}(-\frac{\delta^{ab}%
g^{2}}{2i})(\sigma\varkappa_{q})^{2}(I_{\mu\nu}^{(4)}(k,m_{q},m_{q})+I_{\mu
\nu}^{(4)}(k,m_{q},0)-\nonumber\\
&  \text{ \ \ }-I_{\mu\nu}^{(4)}(k,0,m_{q})-I_{\mu\nu}^{(4)}%
(k,0,0))\nonumber\\
&  =\frac{1}{16\pi^{2}}\frac{2}{\epsilon}\frac{\delta^{ab}g^{2}}{2}\sum
_{q}(\sigma\varkappa_{q})^{2}{\Large (}12\times(+m_{q}^{2})\text{ }g_{\mu\nu
}-\nonumber\\
&  -\frac{1}{3}m_{q}^{2}(k^{2}g_{\mu\nu}-k_{\mu}k_{\nu}){\Large ),}\\
\Pi_{\mu\nu}^{(5\operatorname{div})}(k) &  =\frac{1}{16\pi^{2}}\frac
{2}{\epsilon}\delta^{ab}g^{2}\sum_{q}4\times(\sigma\varkappa_{q})^{2}m_{q}%
^{2}\text{ },
\end{align}

\bigskip in which $P[Q(x)]$ indicates \ the pole part in $x$ of a quantity
$Q(x).$

Thus, the summation of all the divergent contributions leads to
\begin{align}
\sum_{n=1,...,4}\Pi_{\mu\nu}^{(n\operatorname{div})}(k)  &  =-\frac{1}%
{16\pi^{2}}\frac{2}{\epsilon}\frac{\delta^{ab}g^{2}}{6}(k^{2}g_{\mu\nu}%
-k_{\mu}k_{\nu})\sum_{q}(\sigma\varkappa_{q})^{2}m_{q}^{2}\nonumber\\
&  =-\frac{1}{8\pi^{2}}\delta^{ab}g^{2}(k^{2}g_{\mu\nu}-k_{\mu}k_{\nu}%
)\frac{1}{\epsilon}.
\end{align}
This result shows that the sum of all primitively divergent diagrams,
excluding the tadpole term, becomes transverse in the momentum, as gauge
invariance requires. It also arises as being independent of the new six new
parameters $\ \varkappa_{q},$ $q=1,...,6.$. Thus, in this case the one loop
divergence can be eliminated by the counterterms only dependent of the strong
coupling $g$. It can be noted  that tadpole term is being required to be eliminated
by the normal ordering of the  operators in the vertices in order to attain gauge invariance.
This result is consistent  with the  discussions in references \cite{collins,collecot}
 in connection with the compatibility of the normal ordering with the symmetries of the
 physical systems. The same study given  in this section should be done for the various other primitively
divergent diagrams. This task will be considered as a direct extension of the
present work.

\section{Summary}

A proposal of a massless QCD including NJL action terms in a local and
renormalizable form is started to be investigated in detail. The new terms
included in the action determine masses for all the six quarks which are given
by the reciprocal of the new six $flavour$ condensate couplings linked with
each quark type. Here, after defining the divergence indices of the theory and
identifying the primitive divergences, the renormalized Feynman expansion is
presented. The expansion is employed in this work to evaluate the  divergence of
the gluon polarization operator. After the complete evaluation of all the
various primitive divergences, the results will allow to further
consider the one loop renormalization of the model in coming works. The
extension of the study also will be devoted to investigate the possibility to
generate a quark mass hierarchy as a dynamic $flavour$ symmetry breaking in
the context of the model. In this process, it might be suspected that after
the divergences are removed, the contributions to the vacuum energy associated
to diagrams showing two different kinds of fermion lines, might tend to rise
the energy with respect to the diagrams exhibiting equal values of the quark
masses, making them more energetic that the ones in which a single quark mass
parameter gets a finite value. It is interesting to remark that the occurrence
of this flavour symmetry breaking, may be allowed by the fact that included
two-gluon-two-quark vertices directly break chiral invariance. The framework
seems appropriate to dynamically realize the so called Democratic Symmetry
Breaking properties of the mass hierarchy remarked by H. Fritzsch
\cite{fritzsch}. It can be also imagined that the appearance of six different
couplings in the theory, could be reduced to only one by employing the
Zimmermann's reduction of the couplings approach \cite{zimmermann}. This
possibility also suggests a way for linking the model with the SM assuming
that the single coupling could be expected to play the role of the Higgs
field. This property is also suggested by the known results which show that
the Top condensate models can be re-formulated in a way being closer to the SM
\cite{bardeen} thanks in good part to the gauge invariance, which is explicit
in the model. It can be concluded that the discussion supports the starting
idea of the study about that massless QCD could generate an intense
dimensional transmutation effect. Its feasibility will be investigated in the
extension of this work. In ending, it could be helpful to remark that in the
gluodynamic limit of the results of reference \cite{17} (which motivated the
discussion in this paper) the appearance of Gaussian means over color fields
suggested the possibility of a first principles derivation of the linear
confining effects predicted by the stochastic vacuum models of QCD
\cite{doetsch}.

\section*{Acknowledgment}

I would like to express my deep acknowledgements to the: Department of Physics
of the New York University by the kind hospitality during a short visit in
which a starting version of this work was exposed, and the helpful remarks
received during the stay from M. Porrati, G. Gabadadze, A. Reban, J.
Lowenstein, D. Zwanziger and A. Sirlin. Further, I should also deeply
acknowledge the Max Planck Institute for Physics "Werner Heisenberg" (MPI) in
Munich, for the kind invitation to visit this Center during May 2017. At the
MPI, I had the opportunity to finish this work, as well as discussing it with
many colleagues. I very much appreciate the remarks and discussions with D.
Luest, W. Hollik, T. Hahn, E. Seiler, P. Weisz, G. Heinrich, M. Kerner, and S.
Jahn. I should also express my gratitude by the exchanges on the theme
sustained along the time with A. Gonz\'{a}lez, A. Tureanu, M. Chaichian, A.
Klemm, M. Wise, M. Peskin, N. G. Cabo-Bizet and A. Cabo-Bizet. The support
granted by the N-35 OEA Network of the ICTP is also greatly appreciated.
\appendix
\section{\bigskip The set of renormalized vertices}
This appendix is simply devoted to depict the original vertices of the model
in momentum space and their corresponding analytic expressions. They are shown
in the first figure 4 below. The second figure 5 shows the counterterm
vertices in momentum space and their analytic expressions.
\begin{figure}[H]
\begin{center}
\includegraphics[scale=0.35]{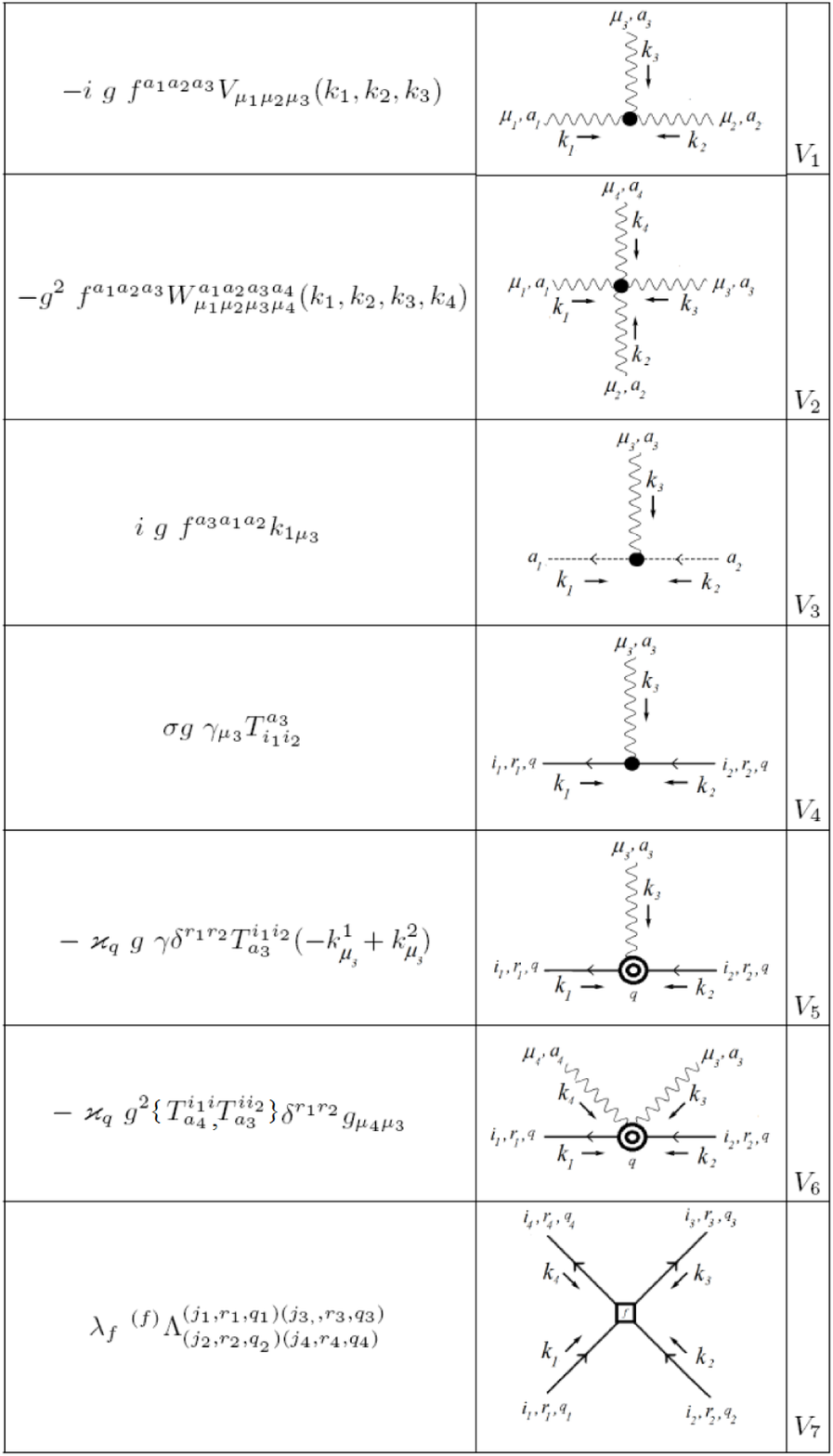}
\end{center}
\caption{The original vertices of the model in terms of the renormalized
fields and couplings. }%
\label{table2}%
\end{figure}

\begin{figure}[H]
\begin{center}
\includegraphics[scale=0.4]{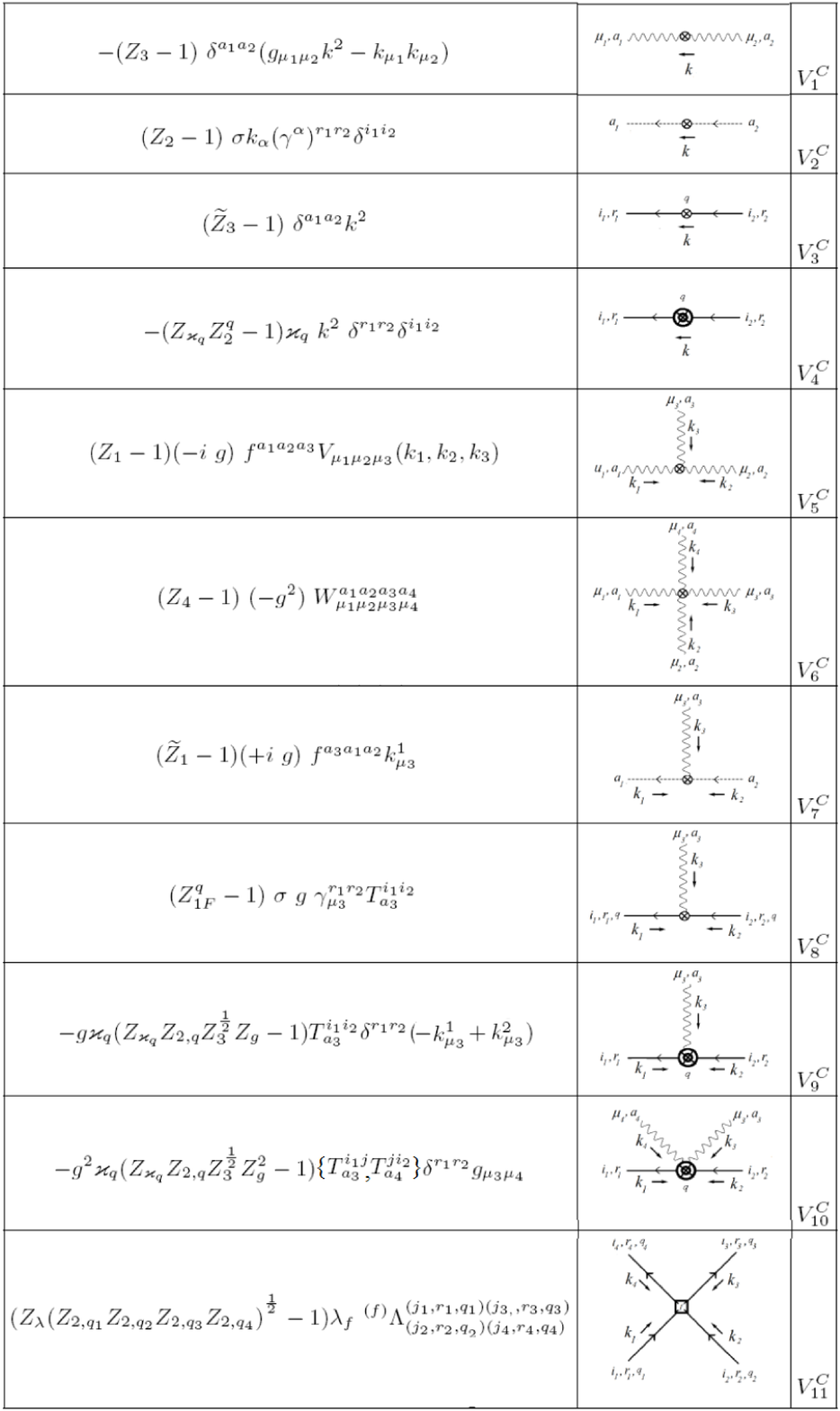}
\end{center}
\caption{The eleven counterterm vertices implementing the renormalization.}%
\label{table3}%
\end{figure}

\end{document}